\documentclass[twocolumn,showpacs,preprintnumbers,amsmath,amssymb]{revtex4}
\usepackage{graphicx}% Include figure files
\usepackage{dcolumn}% Align table columns on decimal point
\usepackage{bm}% bold math

\begin{document}

\preprint{APS/123-QED}

\title{Quantum critical phenomena in a spin-1/2 frustrated square lattice with spatial anisotropy}

\author{H. Yamaguchi$^{1}$, Y. Iwasaki$^{2}$, Y. Kono$^{3}$, T. Okubo$^{4}$, S. Miyamoto$^{1}$, Y. Hosokoshi$^{1}$, A. Matsuo$^{5}$, T. Sakakibara$^{5}$, T. Kida$^{6}$, and M. Hagiwara$^{6}$}
%\author{Second Author}
%\email{yamaguchi@p.s.osakafu-u.ac.jp}
\affiliation{
$^1$Department of Physical Science, Osaka Prefecture University, Osaka 599-8531, Japan\\
$^2$Center for Advanced High Magnetic Field Science (AHMF), Graduate School of Science, Osaka University, Osaka 560-0043, Japan\\
$^3$Department of Physics, Chuo University, Tokyo 112-8551, Japan\\ 
$^4$Department of Physics, the University of Tokyo, Tokyo 113-0033, Japan\\
$^5$Institute for Solid State Physics, the University of Tokyo, Chiba 277-8581, Japan\\
$^6$Center for Advanced High Magnetic Field Science (AHMF), Graduate School of Science, Osaka University, Osaka 560-0043, Japan\\
}

%\affiliation{
Second institution and/or address\\
This line break forced% with \\

\date{\today}% It is always \today, today,

\begin{abstract}
We present a model compound with a spin-1/2 spatially anisotropic frustrated square lattice, in which three antiferromagnetic interactions and one ferromagnetic interaction are competing.  
We observe an unconventional gradual increase in the low-temperature magnetization curve reminiscent of the quantum critical behavior between gapped and gapless phases.
In addition, the specific heat and electron spin resonance signals indicate one-dimensional characteristics.
These results demonstrate quantum critical behavior associated with one-dimensionalization caused by frustrated interactions in the spin-1/2 spatially anisotropic square lattice.
\end{abstract}


\maketitle
%frustrated square latticeから始める、最近の実現例
%強磁性相関を１つ含む場合が本系　⇒　その場合は実効的にS=1三角格子となる
One current key area in condensed matter physics is the search for quantum phenomena associated with magnetic frustration, in which neighboring spins interact via competing exchange interactions that cannot be satisfied simultaneously.
The archetypal example of frustrated magnets is a triangular lattice antiferromagnet. 
The quantum spin liquid picture on the spin-1/2 triangular lattice proposed by Anderson in 1973 has stimulated extensive research on ground states caused by frustration~\cite{RVB}. 
However, subsequent numerical studies have established that its ground state has an antiferromagnetic (AF) order~\cite{120_1,120_2}.
A square lattice can also be made to exhibit frustration via the coexistence of AF and ferromagnetic interactions. 
Such a system has been theoretically studied in association with Josephson-junction arrays~\cite{FFSL_1,FFSL_2,FFSL_3}; however, there have been no reports of ideal model compounds. 
Recently, our spin arrangement design using organic radicals enabled the formation of a spin-1/2 frustrated square lattice with spatial anisotropy.
Such compounds demonstrate that the strong quantum fluctuations and frustrated plaquettes stabilize the quantum states~\cite{TCNQ, PF6}.
The spin model focused on in the present study is one example of such a spin-1/2 spatially anisotropic frustrated square lattice, where three AF interactions and one ferromagnetic interaction form a plaquette, yielding spatial anisotropy.
%Furthermore, since the ferromagnetically coupled spins can form an effective spin-1, this system can be mapped to a spin-1 spatially anisotropic triangular lattice, as shown in Fig. 1(b).

%歪んだ三角格子の理論研究、相図
%1/2の系でone-dimensionality by frustrationと多様な相図を紹介　⇒　S=1の系ではgappedなHaldaneが基底である特異性、そしてHaldaneと秩序の間に臨界点があり予想されているとこと
Frustrated systems have a delicate energy balance; accordingly, additional factors can easily change their ground states. 
One such factor is spatial anisotropy, which is equivalent to lattice distortion.
The spin-1/2 Heisenberg triangular lattice case has been extensively studied.
An anisotropic triangular lattice can be viewed as 1D chains with frustrated interactions between them. 
The frustration effectively decouples the chains, creating a 1D chain system. 
As a consequence, a gapless disordered state analogous to a Tomonaga-Luttinger liquid (TLL) in a 1D system becomes stable in a spin-1/2 anisotropic triangular lattice~\cite{Cs2CuCl4_exp1, Cs2CuCl4_exp2, spin1/2_1, spin1/2_2, spin1/2_3,spin1/2_4}. 
Here we show that a similar frustration-induced dimensional reduction also can occur in certain spin-1/2 anisotropic square lattices as we describe below.

%For the spin-1 case, which corresponds to the mapping model in the present work, a dimensional reduction associated with the spatial anisotropy has also been suggested~\cite{spin1_5,spin1_7, spin1_6,spin1_1}.
%It is a crucial difference that a gapped Haldane state is formed in the spin-1 chain~\cite{haldane} unlike the gapless ground state in the spin-1/2 chain.
%A spiral magnetic phase appears when the interchain frustrated zigzag couplings close the Haldane gap.
%The one-dimensionalization caused by the frustration is predicted to enhance the critical coupling required to close the Haldane gap by an order of magnitude compared to unfrustrated case~\cite{spin1_5,spin1_7, spin1_6, spin1_1, spin1_square1, spin1_square2}. 

%, where 1D chains are formed through $J$ and coupled through frustrating zigzag path $J^{'}$, 

%On the other hand, for the spin-1 case has been little explored in the literature. 

%the Haldane gap is only suppressed at J/J ≈ 0.4 [19], ascompared with J/J ≈ 0.04 for the unfrustrated case [20]. 

%要約
In this letter, we present a model compound with a spin-1/2 spatially anisotropic frustrated square lattice. 
We successfully synthesized single crystals of the verdazyl-based salt [$m$-MePy-V-($p$-F)$_2$]SbF$_6$. 
Our molecular orbital (MO) calculations indicate the formation of a spatially anisotropic square lattice composed of three AF interactions and one ferromagnetic interaction causing frustration.
The low-temperature magnetization curve exhibits an unconventional gradual increase reminiscent of the quantum critical behavior between gapped and gapless phases.
The specific heat and the electron spin resonance (ESR) signals indicate gapless 1D characteristics.
These results demonstrate the quantum critical behavior between gapped 1D-based disordered and two-dimensional (2D) magnetic ordered phases.

%実験方法
We prepared $m$-MePy-V-($p$-F)$_2$ using a conventional procedure~\cite{gosei} and synthesized [$m$-MePy-V-($p$-F)$_2$]SbF$_6$ using a reported procedure for salts with similar chemical structures~\cite{PF6, 3Dhoneycomb}.
Recrystallization using acetonitrile yielded dark-brown [$m$-MePy-V-($p$-F)$_2$]SbF$_6$ crystals.
The crystal structure was determined from intensity data obtained using a Rigaku AFC-8R Mercury CCD RA-Micro7 diffractometer with a Japan Thermal Engineering XR-HR10K. 
High-field magnetization measurements were made using a nondestructive pulse magnet under pulsed magnetic fields of up to approximately 50 T. 
The magnetization curve at 80 mK was measured using a capacitive Faraday magnetometer in a dilution refrigerator.
The magnetic susceptibility was measured using a commercial SQUID magnetometer (MPMS-XL, Quantum Design). 
The experimental result was corrected considering the diamagnetic contributions calculated via Pascal's method. 
The specific heat was measured using a commercial calorimeter (PPMS, Quantum Design) through a thermal relaxation method down to approximately 0.7 K. 
The ESR measurements were performed using a vector network analyzer (ABmm), superconducting magnet (Oxford Instruments), and a laboratory-built cylindrical cavity. 
Considering the isotropic nature of organic radical systems, all the experiments were performed using small randomly oriented single crystals.

%結晶構造
Figure 1(a) shows the molecular structure of [$m$-MePy-V-($p$-F)$_2$]SbF$_6$, which has an $S$ = 1/2 within a molecule.
No indications of structural phase transitions have been observed down to 25 K~\cite{supple1}.
Because this study focuses on low-temperature magnetic properties, we examined the low-temperature crystallographic data at 25 K.
The crystallographic parameters at 25 K are as follows: triclinic, space group $P\bar{1}$, $a$ = 7.134(4) $\rm{\AA}$, $b$ = 11.302(6) $\rm{\AA}$, $c$ = 13.902(7) $\rm{\AA}$, $\alpha$ = 68.83(2) $^{\circ}$, $\beta$ = 82.81(3) $^{\circ}$, $\gamma $ = 85.55(3) $^{\circ}$, $V$ = 1036.4 $\rm{\AA}^3$, and $Z$ = 2. 
We performed MO calculations~\cite{MOcal} to evaluate the exchange interaction between the $S$ = 1/2 spins on the radicals and found four types of dominant interactions~\cite{supple1}, as shown in Fig. 1(b).
They are identified as $J_{1}/k_{\rm{B}}$ = $54.3$ K, $J_{2}/k_{\rm{B}}$
= 21.1 K, $J_{3}/k_{\rm{B}}$ = $-20.8$ K, and $J_{4}/k_{\rm{B}}$ = 18.0 K, which are defined in the Heisenberg spin Hamiltonian given by $\mathcal {H} =
J_{n}{\sum^{}_{\langle i,j \rangle}}\textbf{{\textit
S}}_{i}{\cdot}\textbf{{\textit S}}_{j}$, where $\sum_{ \langle i,j \rangle}$ denotes the sum over the neighboring spin pairs.
As a consequence, $S$ = 1/2 spins form a spatially anisotropic square lattice in the $ab$ plane, in which three AF interactions and one ferromagnetic interaction cause a frustration, as shown in Fig. 1(c).
The SbF$_6$ anions act as spacers between the 2D planes and enhance the 2D characteristics of the present model~\cite{supple1}.
%Assuming that two spins coupled by the ferromagnetic interaction $J_{3}$ is considered as an effective spin $S_{\rm{eff}}$ = 1, this system can be mapped to a spin-1 spatially anisotropic triangular lattice.

%磁化率
Figure 2(a) shows the temperature dependence of the magnetic susceptibility ( $\chi$ = $M/H$) at 0.1 T. 
Above 150 K, it follows the Curie-Weiss law, and the Weiss temperature is estimated to be $\theta_{\rm{W}}$ = $-15.2$(5) K. 
Considering a mean-field approximation with $\theta_{\rm{W}}$, we obtain ($J_{4}$+$J_{2}$+$J_{3}$+$J_{4}$)/$k_{\rm{B}}$ $\sim$ 61 K, which is close to evaluated value using the MO calculation.
A broad peak appears at approximately 30 K, which indicates an AF short-range order in the 2D lattice.
The upturn below approximately 10 K is caused by slight paramagnetic impurities associated with lattice defects; this is commonly observed in verdazyl-based compounds~\cite{PRL,Zn, TCNQ_JPSJ}.
Assuming conventional paramagnetic behavior $C_{\rm{imp}}/T$ and gapless excitations, where $C_{\rm{imp}}$ is the Curie constant of the impurities, the paramagnetic impurities are evaluated as accounting for approximately 2.7 $\%$ of all spins, which is close to the percentage evaluated in a verdazyl-based salt with a similar molecular structure~\cite{3Dhoneycomb}.

%磁化曲線
The magnetization curve exhibits an unusually convex behavior up to approximately 20 T, as shown in Fig. 2(b).
In higher fields, the magnetization increases almost linearly.
Note that this convex behavior becomes more prominent at a lower temperature of 80 mK, which cannot be explained by the contributions of the paramagnetic impurities.
A slight paramagnetic contribution is found in the low-field region of the magnetization at 80 mK, in which paramagnetic impurities are almost fully-polarized above 0.5 T, as shown in Fig. 2(b)  
The trend of the gradual increase is completely different from the concave shape of the magnetization curve observed in general 2D quantum spin systems~\cite{a235Cl3V, tanaka, kagome}.
The overall magnetization behavior is reminiscent of the quantum critical behavior with a square root dependence in the region between gapped singlet and gapless magnetic phases. 
In 1D gapped spin systems, the energy gap disappears when a magnetic field applied, and a quantum phase transition to the TLL phase appears accompanied by a singular square root behavior of the magnetization curve~\cite{TCNQ_JPSJ,kono,okunishi1, okunishi2}.

%比熱
The temperature dependence of the specific heat $C_{\rm{p}}$ is shown in Fig. 3(a). 
There is no peak showing a phase transition to a magnetic order up to 14 T.
In the low-temperature regions below approximately 2 K, where magnetic contributions are expected to be dominant, as observed in other verdazyl-based compounds, $T$-linear behavior ($C_{\rm{p}}/T$ = const.) is observed, as shown in Fig. 3(b). 
This $T$-linear behavior is robust against an applied magnetic field and appears even at 14 T.
Considering the contributions of the low-energy excitations to the specific heat, this $T$-linear behavior indicates the existence of gapless linear dispersions in a 1D AF spin system.
Such 1D characteristics in the dispersion relation are observed in $S$ = 1/2 spatially anisotropic triangular lattices and are confirmed to originate from one-dimensionalization caused by the frustration effect~\cite{Cs2CuCl4_exp1,Cs2CuCl4_exp2}.
Accordingly, the observed 1D characteristics in the present spatially anisotropic square lattice are also expected to originate from one-dimensionalization owing to frustrated interactions.  
Figure 3(c) shows the entropy $S_{\rm{p}}$ obtained via the integration of $C_{\rm{p}}/T$. 
As shown in the following ESR results, the short-range correlation develops remarkably below approximately 15 K. 
Therefore, we evaluated the magnetic entropy $S_{\rm{m}}$ by subtracting the lattice contribution such that $S_{\rm{m}}$ approaches a constant value above approximately 15 K.
We assumed the lattice contribution of the specific heat given by $\alpha$$T^3$, which corresponds to Debye's $T^3$ law for low-temperature regions and has been confirmed to be effective for verdazyl-based compounds. 
We obtained $S_{\rm{m}}$ with an asymptotic behavior toward 15 K using $\alpha$ = 0.013, as shown in Fig. 3(c).
The value of $\alpha$ is consistent with those used for other verdazyl-based compounds~\cite{Zn, a235Cl3V, pBrV, b26Cl2V, 3Br4FV}.
Considering that the entropy change below 15 K is close to the total magnetic entropy of 5.76 ($R$ln2), the magnetic entropy associated with the present $S$ = 1/2 system is almost entirely consumed in the experimental low-temperature regions.

%ESR実験結果
Figure 4(a) shows the temperature dependence of the ESR absorption spectra at 34.13 GHz.
Sharp resonance signals characteristic of organic radical systems are observed, and the changes in the resonance spectra become noticeable with decreasing temperature.
The resonance fields are converted into $g$ $\sim $ 2.0 at all temperatures, which confirms the isotropic nature of the present system.
The temperature dependences of the resonance shift $\Delta$$H_{\rm{res}}$ and the absorption linewidth $\Delta$$H_{\rm{1/2}}$ are shown in Figs. 4(b) and 4(c), respectively. 
The resonance shift $\Delta$$H_{\rm{res}}$ changes dramatically below approximately 15 K, which indicates that the change  in the internal field accompanied by the development of short-range AF correlations becomes remarkable below 15 K.
Correspondingly, the absorption linewidth $\Delta$$H_{\rm{1/2}}$ exhibits a broadening with decreasing temperature.
We fitted the ESR linewidth assuming $\Delta$$H_{\rm{1/2}}$ $\propto$ $T^{-p}$ and obtained $p$ = 0.90(6).
The low-temperature parts of $\Delta$$H_{\rm{1/2}}$ in weakly coupled 1D spin chains are described by a $T^{-1}$ broadening~\cite{furuya, furuya_rei1, furuya_rei2}.
Considering that the evaluated $p$ is close to 1.0, the development of correlations in the present system has 1D characteristics.

%考察
Here, we consider the ground state of the present system.
The experimental results exhibited neither an energy gap nor 2D characteristics, while gapless 1D characteristics were observed.
Furthermore, the nonmonotonic increase in the magnetization curve suggests that the ground state is in the vicinity of a quantum critical point between gapped and gapless phases. 
Considering the observed 1D characteristics, we assumed two types of possible 1D correlations: a $J_{1}$-$J_{2}$ AF alternating chain and a $J_{1}$-$J_{3}$ AF-ferromagnetic alternating chain.
In either case, a singlet ground state formed by the stronger AF interaction $J_{1}$ is separated from a low-lying triplet by an energy gap, which supports the expected quantum critical region near the gapped phase. 
Furthermore, in the later case, the ground state is considered to be equivalent to a Haldane state in the spin-1 AF chain. 
Assuming such a situation, this system can be mapped to a spin-1 spatially anisotropic triangular lattice, where the spins coupled by the ferromagnetic $J_{3}$ form an effective spin-1, as shown in Fig. 1(c).
In spin-1 triangular lattice, a dimensional reduction associated with the spatial anisotropy has been suggested in analogy with the spin-1/2 case~\cite{spin1_5,spin1_7, spin1_6,spin1_1}. 
%The one-dimensionalization is predicted to enhance the critical coupling required to close the Haldane gap by an order of magnitude compared to unfrustrated case~\cite{spin1_5,spin1_7, spin1_6, spin1_1, spin1_square1, spin1_square2}.
The phase boundaries have been evaluated in several numerical studies as a function of $J^{'}/J$, where 1D chains are formed through $J$ and coupled via the frustrated zigzag path $J^{'}$. 
Even though the critical $J^{'}/J$ is still controversial, the phase boundary is predicted to be within the range of 0.3 ${\textless}$ $J^{'}/J$ ${\textless}$ 0.8~\cite{spin1_5,spin1_7, spin1_6, spin1_1}. 
In the present mapped triangular lattice, considering that $J_{2}$ and $J_{4}$ are close values, the ratio $J_{2}/J_{1}$ corresponds to $J^{'}/J$. 
The values of the interactions evaluated from the MO calculations give $J_{2}/J_{1}$ $\sim$ 0.4, which is indeed within the range of the predicted quantum critical point.
Even though it is not clear in which case the actual 1D correlation is, the present work demonstrates the quantum critical behavior between the 1D gapped and 2D gapless magnetic phases, which indicates one-dimensionalization caused by the frustrated interactions in the spin-1/2 spatially anisotropic square lattice

%For the spin-1 case, which corresponds to the mapping model in the present work, a dimensional reduction associated with the spatial anisotropy has also been suggested~\cite{spin1_5,spin1_7, spin1_6,spin1_1}.
%It is a crucial difference that a gapped Haldane state is formed in the spin-1 chain~\cite{haldane} unlike the gapless ground state in the spin-1/2 chain.
%A spiral magnetic phase appears when the interchain frustrated zigzag couplings close the Haldane gap.
%The one-dimensionalization caused by the frustration is predicted to enhance the critical coupling required to close the Haldane gap by an order of magnitude compared to unfrustrated case~\cite{spin1_5,spin1_7, spin1_6, spin1_1, spin1_square1, spin1_square2}. 

%For the spin-1 case, which corresponds to the mapping model in the present work, a dimensional reduction associated with the spatial anisotropy has also been suggested~\cite{spin1_5,spin1_7, spin1_6,spin1_1}.
%It is a crucial difference that a gapped Haldane state is formed in the spin-1 chain~\cite{haldane} unlike the gapless ground state in the spin-1/2 chain.
%A spiral magnetic phase appears when the interchain frustrated zigzag couplings close the Haldane gap.
%The one-dimensionalization caused by the frustration is predicted to enhance the critical coupling required to close the Haldane gap by an order of magnitude compared to unfrustrated case~\cite{spin1_5,spin1_7, spin1_6, spin1_1, spin1_square1, spin1_square2}. 

%結論
In summary, we succeeded in synthesizing single crystals of the verdazyl-based salt [$m$-MePy-V-($p$-F)$_2$]SbF$_6$. 
The MO calculations indicated that the $m$-MePy-V-($p$-F)$_2$ molecules formed an $S$ = 1/2 spatially anisotropic frustrated square lattice, in which three AF interactions and one ferromagnetic interaction are competing. 
The low-temperature magnetization curve exhibited an unconventional gradual increase reminiscent of the quantum critical behavior between gapped and gapless phases.
Furthermore, the specific heat and the ESR signals indicated gapless 1D characteristics.
These properties can be understood as the quantum critical behavior between the gapped 1D-based disordered and gapless 2D magnetic ordered phases.    
Therefore, we demonstrated that the model compound of the $S$=1/2 spatially anisotropic frustrated square lattice, [$m$-MePy-V-($p$-F)$_2$]SbF$_6$, realizes quantum critical behavior associated with one-dimensionalization caused by frustrated interactions.
The present material provides insights into the ground states realized in frustrated square lattices, which have not been fully explored, and will stimulate further studies on quantum phenomena associated with magnetic frustration in condensed matter physics.

\begin{figure}[t]
\begin{center}
\includegraphics[width=20pc]{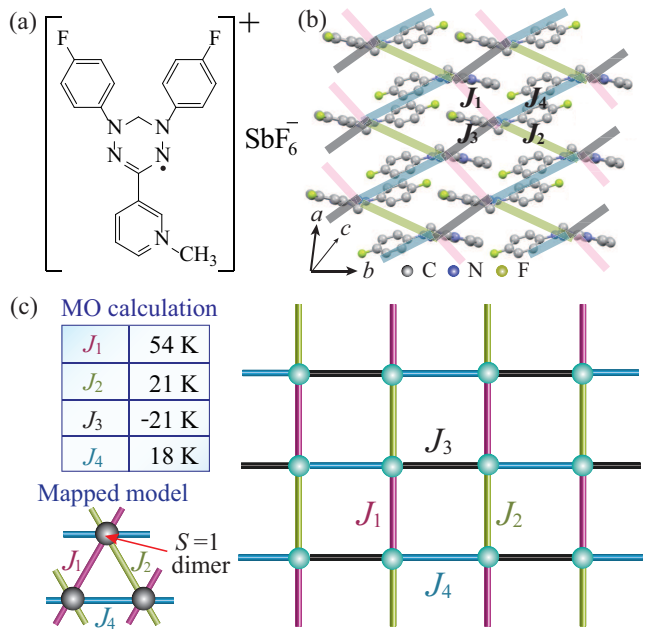}
\caption{(color online) (c) Molecular structure of [$m$-MePy-V-($p$-F)$_2$]SbF$_6$. (d) Crystal structure forming the spin model in the $ab$ plane. Hydrogen atoms are omitted for clarity. (c) Spin-1/2 spatially anisotropic frustrated square lattice composed of three AF interactions, $J_{1}$, $J_{2}$, and $J_{4}$, and one ferromagnetic interaction $J_{3}$ in [$m$-MePy-V-($p$-F)$_2$]SbF$_6$. The mapped model corresponds to a spin-1 spatially anisotropic triangular lattice composed of the spins dimerized by $J_{3}$.}
\end{center}
\end{figure}

\begin{figure}[t]
\begin{center}
\includegraphics[width=18pc]{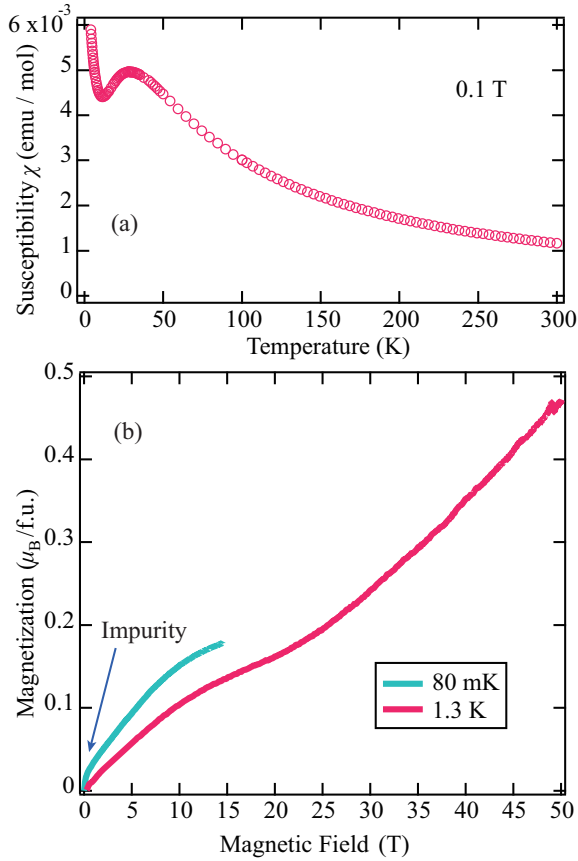}
\caption{(color online) (a) Temperature dependence of the magnetic susceptibility ($\chi$ = $M/H$) of [$m$-MePy-V-($p$-F)$_2$]SbF$_6$ at 0.1 T . 
(b) Magnetization curves of [$m$-MePy-V-($p$-F)$_2$]SbF$_6$ at 80 mK and 1.3 K. The arrow shows the contribution of the paramagnetic impurities.}\label{f2}
\end{center}
\end{figure}

\begin{figure}[t]
\begin{center}
\includegraphics[width=20pc]{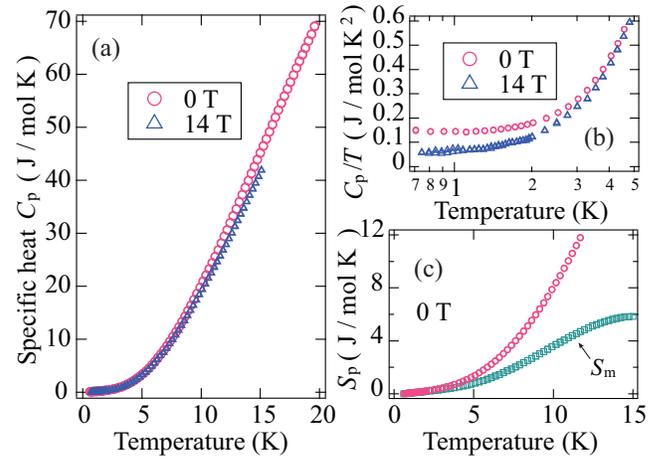}
\caption{(color online) (a) Temperature dependence of the specific heat $C_{\rm{p}}$ of [$m$-MePy-V-($p$-F)$_2$]SbF$_6$ at 0 and 14 T. (c) Low-temperature region of $C_{\rm{p}}/T$. 
(b) Entropy $S_{\rm{p}}$ obtained via the integration of $C_{\rm{p}}/T$ at zero-field. $S_{\rm{m}}$ is the magnetic entropy evaluated by subtracting the lattice contribution. 
}\label{f3}
\end{center}
\end{figure}

\begin{figure}[t]
\begin{center}
\includegraphics[width=20pc]{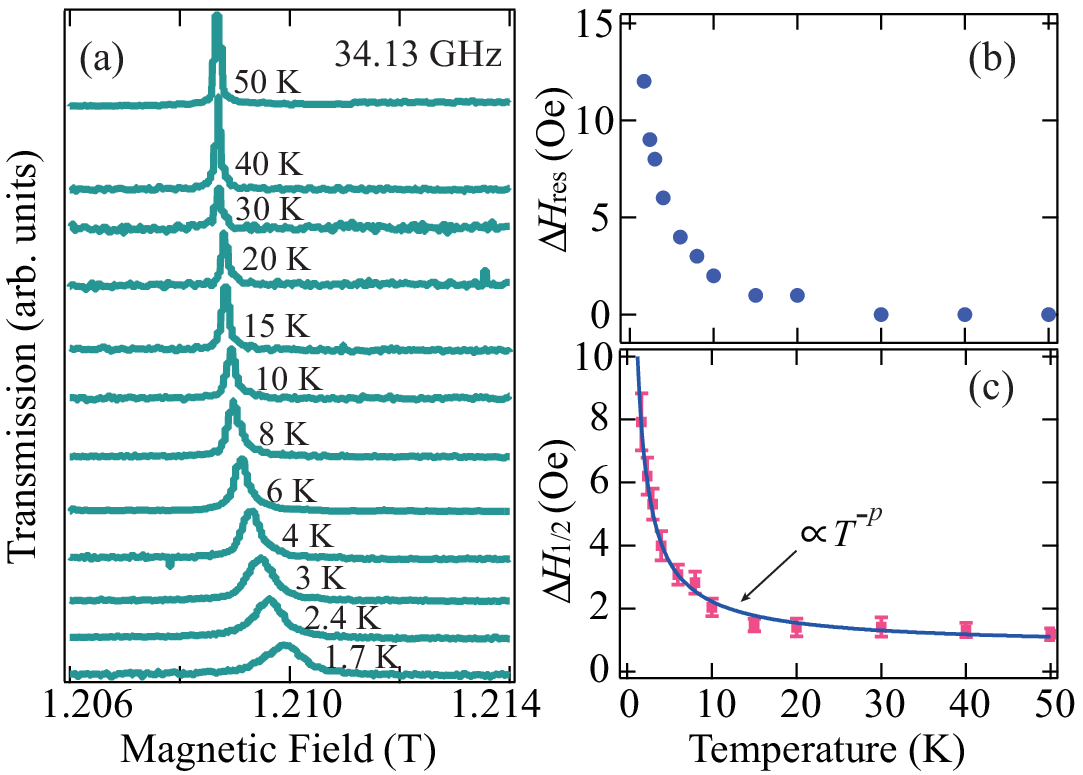}
\caption{(color online) (a) Temperature dependence of the ESR absorption spectra of [$m$-MePy-V-($p$-F)$_2$]SbF$_6$ at 34.13 GHz. Temperature dependence of the (b) resonance field shift and (c) linewidth (full width at half maximum) obtained from the ESR spectra. The solid line indicates the $T^{-p}$ fit. }\label{f4}
\end{center}
\end{figure}

\begin{acknowledgments}
We thank S. C. Furuya for valuable discussion.
This research was partly supported by Grant for Basic Science Research Projects from KAKENHI (No. 19J01004 and No. 19H04550), the Murata Science Foundation, Kyoto Technoscience Center, SEI Group CSR Foundation, and Nanotechnology Platform Program (Molecule and Material Synthesis) of the Ministry of Education, Culture, Sports, Science and Technology (MEXT), Japan. 
A part of this work was performed at the Center for Advanced High Magnetic Field Science in Osaka University under the interuniversity cooperative research program of the joint-research program of ISSP, the University of Tokyo.
\end{acknowledgments}

%%%%%%%%%%%%%%%%%%%%%%%%%%%%%%%%%%%%%%%%%%%%%%%%%%%%%%%%%%%%%%
%%%%%%%%%%%%


\begin{thebibliography}{99}
\bibitem{MOcal}
The MO calculations were performed using the UB3LYP method in the Gaussian 09 program package. 
The basis set is 6-31G. 
For the estimation of intermolecular magnetic interaction, we applied the previously reported evaluation scheme~\cite{MO_method}.
\end{thebibliography}
\end{document}